\documentclass[aps,pre,preprint,amsmath,amssymb,superscriptaddress,floatfix,showpacs]{revtex4}
\usepackage[dvips]{graphicx}
\usepackage{latexsym}
\usepackage{revsymb}
\usepackage{amsfonts}
\usepackage{amsmath}
\usepackage{amssymb}
\usepackage{bm}

\begin{document}


\title{On the Steady State Distributions for Turbulence}

\renewcommand{\thefootnote}{\fnsymbol{footnote}}

\author{Djordje Minic}
\affiliation{Department
of Physics, Virginia Tech, Blacksburg, VA 24061, USA}

\author{Michel Pleimling}
\affiliation{Department
of Physics, Virginia Tech, Blacksburg, VA 24061, USA}

\author{Anne E. Staples}
\affiliation{Department of Engineering Science and Mechanics, Virginia Tech, Blacksburg, VA 24061, USA}
%
%
%

\date{\today}
\begin{abstract}
We propose explicit forms for the steady state distributions governing
fully developed turbulence in two and three spatial dimensions. 
We base our proposals on the crucial importance of the
area and volume preserving diffeomorphisms in the space of velocities.
We argue that these distributions can lead to the relevant (Kolmogorov and Kraichnan)
scaling laws.

\end{abstract}
\pacs{47.27.Ak,47.27.Gs,11.15.-q}

\maketitle
\pagestyle{plain}

In this note we propose universal steady state distributions for fully
developed turbulent flows in two and three dimensions (2d and 3d)
building on the recent work which aims to relate modern quantum field theory, string theory
and fluid dynamics \cite{turb}.
The basic dynamical equation for turbulent flow is the non-linear Navier-Stokes equation (we use the sum convention throughout the paper)
\begin{equation}
\rho (\partial_t v_i + v_j \partial_j v_i) = - \partial_i p + \nu \partial_j^2 v_i,
\end{equation}
with the incompressibility condition $\partial_i v_i = 0$, where $v_i$ is a component of the velocity field of the flow, $p$ is pressure, and $\rho$ is the fluid density \cite{review}.
In this note we are interested in fully developed
turbulence, or turbulence in the limit of infinite Reynolds number $R$. As $R = L v/{\nu}$
goes to infinity, with $v = \sqrt{v_i v_i}$, whereas $L$ is a characteristic scale and $\nu$ is the kinematic viscosity, we effectively have the limit of vanishing
viscosity $\nu \to 0$. In this regime, all the various possible symmetries are restored in a statistical sense,
calling for a probabilistic description of what is in essence a deterministic system (strongly dependent on the boundary conditions). Therefore,
in computing correlators of the velocity field, we should use the statistical and
quantum field theoretic descriptions of turbulence \cite{qft}.

In particular we are interested in the form of a steady state turbulent distribution $P(\vec{v})$
which should imply the various famous scaling laws (the Komogorov scaling in 3d \cite{ko} and 
both Kolmogorov and Kraichnan scaling laws in 2d \cite{kr}). The two-point correlators determined
by these scalings should be computable from $P(\vec{v})$ as follows
\begin{equation}
\langle v_i(l) v_j (0) \rangle \equiv 
\frac{\int D \vec{v} v_i(l) v_j(0) P(\vec{v})}{\int D \vec{v} P(\vec{v}) }.
\end{equation}
The fundamental question is: what is $P(\vec{v})$?
In this note we wish to argue that the answer to this question is:
\begin{equation}
P(\vec{v}) = \exp[-S_{K} (\vec{v})],
\end{equation}
where we assume that the expression for the $S_K(\vec{v})$ is local and universal (see below). Both of these 
assumptions might be challenged:
we can a priori expect non-local factors in $S_K(\vec{v})$. 
Also, the relative locations of the Kolmogorov and Kraichnan distributions in 2d turbulence energy spectra
depend on the (location of the) forcing, which challenges universality.
We also note that the notion of universality might be challenged by an a priori dependence
on the boundary conditions. We will simply assume that universal distributions exist
in the rest of this note, in spite of these caveats.
Finally, we note that if universality applies to turbulence, it is expected at short 
distance \cite{qft}, as opposed to the usual long distance universality associated with
quantum field theory and critical phenomena from equilibrium physics.

The central message of this note is that
$S_K$ is what we call the Kolmogorov distribution in 3d,
determined by the effective action for a 3d gauge theory
based on volume preserving diffeomorphisms. Alternatively,
$S_{K}$ is what we call the Kraichnan distribution in 2d, 
determined by the effective action for a 2d gauge theory based on area preserving diffeomorphisms.
In both cases the diffeomorphisms act in velocity space.
In both cases we crucially use the Galilean symmetry in the space of velocities, so that
the effective respective actions involve only derivatives in $v_i$.

We pause briefly to remark that in order to fully characterize systems far from equilibrium
knowledge of the probability currents is needed in addition to the steady state probability 
distributions \cite{zia}. However, a discussion of these currents is outside of the scope
of the present paper.

Even though the puzzle of fully developed turbulence is essentially a strongly coupled
problem, we motivate our discussion (and ultimately, our proposals) by some rather elementary observations at
weak coupling.
We start with 2d and then move on to 3d.
In 2d the Lagrangian  description of the fluid is generated by 
the following Lagrangian \cite{lenny} (in the notation of \cite{lenny})
\begin{equation}
L_2= \int d^2y \rho_0 [ \frac{m}{2} \dot{x}^2 - V( \rho_0 |\frac{ \partial y}{\partial x}|)].
\end{equation}
where $\rho_0$ is the constant density in the co-moving coordinates
and the real space density is $\rho_0 |\frac{ \partial y}{\partial x}|$,
where $|\frac{ \partial y}{\partial x}|$ denotes the Jacobian connecting the 
co-moving coordinates $y$ and the continuum fields $x_i(y, t)$.
The Lagrangian $L_2$ is invariant under area preserving diffeomorphisms in 2d \cite{lenny}:
\begin{equation}
y_i' = y_i +f_i(y) \quad;\quad \delta x_a = \frac{\partial x_a}{\partial y_i} f_i(y),
\end{equation}
where for area preserving diffeomorphisms ($\epsilon_{ij}$ being the Levi-Civita symbol in 2d)
\begin{equation}
f_i = \epsilon_{ij} \frac{\partial \Lambda(y)}{\partial y_j},
\end{equation}
and where $\Lambda(y)$ is an arbitrary function, which generates these ``gauge''
transformations. This equation in turn
leads to the 2d Poisson bracket action:
\begin{equation}
\delta x_a = \frac{\partial x_a}{\partial y_i}\epsilon_{ij} \frac{\partial \Lambda(y)}{\partial y_j}
\equiv \{\Lambda, x_a\}~.
\end{equation}
Suppose we consider small motions of the 2d fluid \cite{lenny}:
\begin{equation}
x_i = y_i + e \epsilon_{ij} A_j,
\end{equation}
where $e$ is the small coupling (and where we have absorbed the factor of $\rho_0$ compared to
\cite{lenny}). To linear order the area preserving diffeomorphisms become the usual gauge transformations
\begin{equation}
\delta A_i = \partial_i \Lambda,
\end{equation}
and the quadratic Lagrangian that is invariant under this transformation is
just the usual Maxwell Lagrangian
\begin{equation}
L_2 \sim \frac{1}{2 e^2} \int d^2y [ \dot{A_i}^2 - (\nabla \times \vec{A})^2].
\end{equation}
The inclusion of non-linear terms can be done by extending the linear gauge transformations
to their non-Abelian completion. This leads to a non-Abelian gauge theory,
with the following map between the full area preserving diffeomorphism group and the non-Abelian transformations
generated by a commutator of two matrices
\begin{equation}
\{\Lambda, x\} \to [\lambda, X].
\end{equation}
Here we have used the standard \cite{apd} mapping between
the Poisson brackets and commutators of infinite square matrices $\lambda$ and $X$.
The corresponding gauge theory action reads as
\begin{equation} \label{2dYM}
S_A=  \frac{1}{2 e^2} \int d^2 y [  \dot{A_i}^2 - (\partial_i A_j - \partial_j A_i + [A_i, A_j])^2].
\end{equation}
We will return to this gauge theory action in what follows.
In 3d we can extend the presentation of \cite{lenny} by considering
\begin{equation}
L_3= \int d^3y \rho_0 [ \frac{m}{2} \dot{x}^2 - V( \rho_0 |\frac{ \partial y}{\partial x}|)].
\end{equation}
The Lagrangian $L_3$ is invariant under volume preserving diffeomorphisms
\begin{equation}
\delta x_a = \frac{\partial x_a}{\partial y_i}\epsilon_{ijk} \frac{\partial \Lambda_1(y)}{\partial y_j} 
\frac{\partial \Lambda_2(y)}{\partial y_k}.
\end{equation}
where $\Lambda_1$ and $\Lambda_2$ are the generators of the volume preserving
gauge transformations.
This is equivalent to the following Nambu bracket \cite{nambu}:
\begin{equation}
\delta x_a \equiv \{\Lambda_1, \Lambda_2, x_a\},
\end{equation}
where, by definition 
\begin{equation}
\{A, B, C\}\equiv \epsilon_{abc} \partial_a A \partial_b B \partial_c C.
\end{equation}
Here $A,B,C$ are three functions of three spatial coordinates $x,y,z$.
This classical bracket seems to be naturally generalized to a triple algebraic structure \cite{nambu, triples, vector}
\begin{equation}
[A_i, A_j, A_k] \equiv \epsilon_{abc} A_a A_b A_c
\end{equation}
Suppose we consider small motions of the 3d fluid in analogy with the 2d case:
\begin{equation}
x_i = y_i + f \epsilon_{ijk} B_{jk},
\end{equation}
where $B_{jk}= - B_{kj}$ and $f$ denotes a small coupling.
The ``quantization'' of the volume preserving diffeomorphisms is a 
more involved problem \cite{nambu, triples}. Nevertheless, even in this case
we expect
a mapping between the full volume preserving diffeomorphism group and the 3-bracket
\begin{equation}
\{\Lambda_1, \Lambda_2, x\} \to [\lambda_1, \lambda_2, X],
\end{equation}
where $\lambda_1$ and $\lambda_2$ are the appropriate matrix realizations of
$\Lambda_1$ and $\Lambda_2$ \cite{triples}.
The 3d action invariant under the linear part of this transformation is
\begin{equation}
S_B = \frac{1}{2 f^2}  \int d^3y [  \dot{B}_{ij}^2 - (\partial_i B_{jk} +\partial_j B_{ki} +\partial_k B_{ij})^2].
\end{equation}
Note that the explicit linear volume preserving transformations are
\begin{equation}
\delta B_{ij} = \partial_i \Lambda_1 \partial_j \Lambda_2 - \partial_i \Lambda_2
\partial_j \Lambda_1,
\end{equation}
where $B_{ij}$  is dual to a 3-vector in three dimensions,
$
a_i = \frac{1}{2} \epsilon_{ijk} B_{jk}
$.

The above linear analyses seem removed from such
a strongly coupled problem as fully developed turbulence \cite{zakharov}. 
Still the linear analysis is useful, because in the stationary case we can formally replace
$x_i \to v_i$ and talk about velocity Lagrangians.
That the corresponding Lagrangians should be given in terms of the derivatives of velocity is
fixed by the Galilean symmetry, $v_i \to v_i + u_i$, for a fixed velocity with component $u_i$.
Thus, if we follow the linear analysis in the velocity space and 
drop the time dependence because of our interest in the stationary
distributions, in 2d we have the following natural theory
that is consistent with area preserving diffeomorphisms involving $v_i$:
\begin{equation}
S^{(2)}_K=  \frac{1}{2 g^2} \int d^2 y [  (\partial_i v_j - \partial_j v_i + \{v_i, v_j\})^2],
\end{equation}
where $g^2$ denotes the appropriate dimensionful coupling constant and $i,j=1,2$ (the factor $ \frac{1}{2 g^2}$ is canonical).
This theory is invariant under 
\begin{equation}
\delta v_i = \{ F, v_i \},
\end{equation}
where $F$ generates area preserving gauge transformations in $v_i$ space.
We want to argue that this is the stationary distribution we have been
looking for in 2d.
First, we attempt to justify this guess on more general grounds:
Obviously we have the Galilean invariance in the inertial range
$v_i \to v_i + u_i$ (where $u_i$ is constant)
because $S^{(2)}_K$ is a functional of the derivatives of $v_i$.
Second, it is reasonable to expect that $S^{(2)}_K$ is governed by the conserved quantities.
(Recall the case of equilibrium statistical mechanics, i.e. the Boltzmann-Gibbs distribution, where $S$ is simply the energy, an
additive conserved quantity).
In our situation we have vorticity ($\vec{\omega} = \nabla \times \vec{v}$) squared and velocity squared, as the natural
conserved quantities \cite{review, ko, kr}.
Therefore, if we start with the vorticity squared ($\omega^2$) term we see that this is
really the quadratic part of our guess for $S^{(2)}_K$. This term is invariant under
the linear gauge transformations in the space of velocities.
However, by going to real space, we may invoke the full non-linear
group of area preserving coordinate transformations generated by
the Jacobian (the Poisson bracket) in 2d space. Then by concentrating on the steady state
regime we may claim the same symmetry in the $v$ space which would
lead us to the above proposal.

This action should be compared to the 2d Yang-Mills theory action $S_A$ given in equation (\ref{2dYM}).
We can use the covariant derivative
$
\partial_i \to D_i \equiv \partial_i + A_i,
$
to rewrite the usual 2d Yang-Mills theory Lagrangian
as
$[D_i, D_j]^2 \equiv (\partial_i A_j - \partial_j A_i + [A_i, A_j])^2$.
This procedure can be immediately generalized to 3d so that we have
a theory consistent with volume preserving diffeomorphisms in the space of $v_i$.
By using the covariant derivative
$\partial_i \to D^v_i \equiv \partial_i + v_i$,
we can immediately rewrite the above guess for the 2d action 
\begin{equation}
S^{(2)}_K=  \frac{1}{2 g^2} \int d^3 y [  \{D^v_i, D^v_j\}^2],
\end{equation}
and then extrapolate our 2d proposal to the natural proposal for the Kolmogorov distribution in 3d
\begin{equation}
S^{(3)}_K=  \frac{1}{2 \tilde{g}^2} \int d^3 y [  \{D^v_i, D^v_j, D^v_k\}^2].
\end{equation}
Here $\tilde{g}^2$ denotes the appropriate dimensionful coupling constant.
The crucial non-linear part
which replaces $\int d^2 y (\{v_i, v_j\}^2)$ is
\begin{equation}
S^{(3)}_K \sim \frac{1}{2 \tilde{g}^2} \int d^3 y [  \{v_i, v_j, v_k\}^2].
\end{equation}
This action is fixed now by the volume preserving diffeomorphisms in the 3d velocity space
\begin{equation}
\delta v_i = \{ F_1, F_2, v_i \},
\end{equation}
where $F_1$ and $F_2$ generate the volume preserving gauge transformations.
The argument which leads to this proposal is just the repetition of the argument we have
presented for the 2d Kraichnan distribution. 

These are thus our explicit proposals for the turbulent distributions in
2d and 3d: they are encoded in the expressions for $S^{(2)}_K$ and $S^{(3)}_K$.
Do these educated guesses give the correct results?
We concentrate on the case of 2d turbulence.
Let us remember that for the Kraichnan scaling in 2d we
want to derive
$
\langle v_i(l) v_j(0)\rangle \sim l^2 \delta_{ij},
$
and for the Kolmogorov scaling law
$
\langle v_i(l) v_j(0)\rangle \sim l^{\frac{2}{3}} \delta_{ij}.
$
In order to discuss the validity of our proposal, we turn to the
natural loop variables introduced by Migdal \cite{migdal}.
The natural loop variable, the Migdal loop \cite{migdal}
\begin{equation}
W_M (C)\equiv \langle \exp(-\frac{1}{\nu} \int_C v_i dx_i) \rangle,
\end{equation}
(where $C$ is a contour and the viscosity $\nu$ plays the role
of an effective $\hbar$ \cite{migdal, turb}) 
allows us to rewrite the Naiver-Stokes equations \cite{migdal}
as an effective Schrodinger equation 
\begin{equation}
i \nu \partial_t W(C) = H_C W(C),
\end{equation}
with the appropriate loop equation Hamiltonian $H_C$ \cite{migdal}.
In 3d Migdal observed a self-consistent scaling solution of this equation
in the $\nu \to 0$ limit (a WKB limit in this problem)
$
W_M(C) \sim \exp( - [\frac{A}{A_0}]^{\frac{2}{3}}),
$
which precisely corresponds to the Kolmogorov scaling.
In 2d, the Kraichnan scaling leads to
the area law \cite{turb} for the Migdal loop
$
W_M(C) \sim \exp( -A/A_0).
$
In the
above gauge theory of 2d velocities this area law is very natural,
because of the fact that in the corresponding 2d Yang-Mills theory, the Wilson loop (the natural
analog of the Migdal loop)
\begin{equation}
W(C)\equiv \langle \exp(-\int_C A_i dx_i) \rangle,
\end{equation}
obeys the same area law \cite{wilson}
$
W(C) \sim \exp(-A/A_0).
$
Given the precise structural mapping between the 2d Kraichnan theory 
(defined by $S^{(2)}_K$) and
the 2d gauge theory, this area law scaling for the loop variables should be obeyed, and the Kraichnan scaling
should thus follow from our proposed steady state distribution.
This is one of the central observations of this note.

To summarize:
Given the above analogy of the proposal for the Kraichnan distribution and
what we know about the 2d Yang-Mills theory,
Kraichnan's scaling implies the area law for the Migdal loop operator.
In view of this, and assuming locality and universality, our proposal for the Kraichnan distribution, viewed as the gauge theory of 2d 
area preserving diffeomorphisms in velocity space, is essentially unique.

How about the Kolmogorov scaling in 2d? Does this scaling also
follows from our proposal for the universal steady state distribution of 2d turbulence?
Here we can only offer a very heuristic argument. Preliminary numerical
investigations \cite{staples} seem to indicate that the Kolmogorov scaling does
agree with our proposal.

We start our intuitive argument regarding the Kolmogorov scaling in 2d by
noting that in the small momentum regime
the linear term in our 2d gauge theory in velocity space dominates and this is where we expect Kraichnan's scaling.
In the high momentum regime the Jacobian term (the Poisson bracket) dominates. In some
sense this is an effective ``mass term'' (or the restoring force quadratic term from the harmonic motion) that we could associate with
the kinetic energy and the velocity squared term. Given the fact that the energy cascade leads to the Kolmogorov scaling \cite{review}
we should thus expect the Kolmogorov scaling at high momenta.
Thus, naively we expect Kraichnan's scaling at low momenta and 
Kolmogorov's scaling at high momenta \cite{foot1}.
(The issue of the possible dependence on forcing in the mid momentum range makes this intuition in principle more complicated \cite{review}.) 
Nevertheless, by continuing with this picture, we could envision a large Migdal 
loop and apply the area law there (in the large distance, low momentum limit).
In the center of this large loop we envision a much smaller loop, separated by some
radial scale from the large one. The area in between the large and the small loops
sweeps an effective volume as the large loop sweeps its area.
We can then dimensionally translate the area $A$ of the surface swept by the large loop
into this effective volume $V$, $A = V^{2/3}$.
This effective volume is in turn the area of the smaller loop times the distance between the scales.
When we shrink this radial distance to one, we get the area
of the smaller loop (which we might call the Kolmogorov loop) to the power 2/3, i.e. $A^{2/3}$, which
would correspond to Kolmogorov's scaling of the Migdal loop (i.e. $\exp(-(A/A_0)^{2/3})$).
Needless to say, this is a very heuristic picture.
The big caveat in this obviously qualitative argument is the issue of forcing in the mid momentum range.
This intuitive argument has been tested in the preliminary numerical simulations \cite{staples}.
The results are qualitatively encouraging and will be discussed elsewhere \cite{staples}.

There exists a natural reduction from 3d to 2d when considering the reduction
from volume preserving diffeomorphisms to area preserving ones
\begin{equation}
\{\Lambda_1, \Lambda_2, x_a\} \to \{\Lambda_1, x_a\},
\end{equation}
(provided $\Lambda_2 = 1$) and the corresponding reduction of the
3-bracket to the 2-bracket
\begin{equation}
[\lambda_1, \lambda_2, X] \to [\lambda_1, X],
\end{equation}
(provided $\lambda_2 =1$).
Given the above qualitative argument concerning the emergence of the Kolmogorov
scaling (via the $\exp(-(A/A_0)^{2/3})$ scaling of the Migdal loop) and this reduction procedure
we would expect that our proposal for the 3d steady state distribution leads immediately to
the Kolmogorov scaling. Of course, only numerical tests (which are considerably more difficult in 3d as compared to 2d) could confirm this expectation.
Given this reduction we can also propose an interpolation between the 2d and 3d steady state distributions
for quasi-2d fluids of finite thickness by looking at
\begin{equation}
 S^{(2,3)}_K = \frac{1}{2 g^2} \int d^2 y [  (\partial_i v_j - \partial_j v_i + \{v_i, v_j\})^2]
+  \frac{1}{2 \tilde{g}^2} \int d^3 y [  \{v_i, v_j, v_k\})^2].
\end{equation}
Here one should perform the expansion of the 3-bracket in the third anisotropic direction
\begin{equation}
\{A, B, C\}\equiv \dot{A} \{B,C\} + \dot{B}\{C,A\} +\dot{C} \{A,B\},
\end{equation}
where the dot derivative is with respect to the anisotropic third direction.

Finally, we note another general caveat:
Presumably turbulent laws are generated by some boundary conditions
(or by stirring) that are responsible for the cascades
and the ``energy anomaly'' in the Kolmogorov case \cite{review, qft}.
How do we see this in our proposal? We should remember that we need
to compute the functional integral over the velocities ($\int D\vec{v} \exp(-S_{K})$) and
the measure in that functional integral might transform
anomalously once we start worrying about the boundary
conditions and not only the symmetries. That is, the velocity space area preserving diffeomorphisms in 2d
and volume preserving diffeomorphisms in 3d. 
We expect that our discussion is correct as long as boundary effects are neglected.

In conclusion, we have proposed universal statistical distributions for
fully developed turbulence in 2d and 3d.
The proposed steady state distributions for turbulence in 2d and 3d are essentially
fixed by the Galilean invariance and the area and volume preserving gauge transformations,
in 2d and 3d, respectively. We have argued that the Kraichnan scaling in 2d should follow
based on what is known about the non-Abelian gauge theory in 2d.
We have also argued that the Kolmogorov scaling should be consistent with
our proposal, based on a very heuristic picture.
Obviously, only a thorough numerical study can show whether our proposals
are true or not. We will address these numerical issues in an upcoming publication.

{\bf Acknowledgements:}
We would
like to thank Martin Kruczenski, Jack Ng, Chia Tze and Erich Poppitz for
interesting discussions. 
D.M. thanks the Galileo Galilei Institute for Theoretical Physics in Florence for the hospitality and the INFN for partial support during the completion of this work.
D.M.
is supported in part by the US Department of Energy
under contract DE-FG05-92ER40677. 
M.P. is supported by the US National
Science Foundation through DMR-0904999.



\begin{thebibliography}{100}

\bibitem{turb}
Vishnu Jejjala, Djordje Minic, Y. Jack Ng, and Chia-Hsiung Tze,
\ Class.\ Quant.\ Grav. {\bf 25}, 225012 (2008) [arXiv:0806.0030];
\ Mod. \ Phys. \ A {\bf 25}, 2541 (2010) [arXiv:0912.2725];
\ Int. \ J. \ Mod. \ Phys, {\bf D19}, 2311(2010) [arXiv:1005.3254].

\bibitem{review}
L. D. Landau and E. M. Lifshitz, {\it Fluid Dynamics}, Oxford: Pergamon Press (1987);
U. Frisch, {\it Turbulence}, Cambridge: University Press (1995); A. Monin and A. Yaglom,
{\it Statistical Fluid Mechanics}, Cambridge: MIT Press (1975).

\bibitem{qft}
See, for example,
K. Gawedzki, arXiv:hep-th/9710187, or,
A. M. Polyakov, Nucl. Phys. {\bf B396}, 367 (1993) [arXiv:hep-th/9212145];
arXiv:hep-th/9506189,
and references therein.

\bibitem{ko}
A. N. Kolmogorov, Dokl. Akad. Nauk SSSR {\bf 30}, 299 (1941).

\bibitem{kr}
R. H. Kraichnan, Phys. Fluids {\bf 10}, 1417 (1967); R. H. Kraichnan and D. Montgomery,
Rep. Prog. Phys. {\bf 43}, 547 (1980).

\bibitem{zia} R. K. P. Zia and B. Schmittmann, J. Phys. {\bf A39}, L409 (2006);
R. K. P. Zia and B. Schmittmann, J. Stat. Mech. {\bf (2007)} P07012.

\bibitem{lenny}
See L. Susskind, {\it The Quantum Hall Fluid and Non-Commutative Chern-Simons Theory},
 hep-th/0101029, for a nice intuitive discussion.
 
 \bibitem{apd}
 J. Goldstone, unpublished; J. Hoppe, MIT Ph.D. thesis, 1982 and in
{\it Proc. Int. Workshop on ConstraintÕs Theory and Relativistic Dynamics},
G. Longhi and L. Lusanna, eds. (World Scientific, 1987); J. Hoppe,
Int. J. Mod. Phys. {\bf A4} (1989) 5235; D. Fairlie, P. Fletcher and C. Zachos,
J. Math. Phys. {\bf 31} (1990) 1088.


\bibitem{nambu}
Y. Nambu, Phys. Rev. {\bf D7}, 2405 (1973);
Also, 
 L.~Takhtajan,
  Commun.\ Math.\ Phys.\  {\bf 160}, 295 (1994)
  [arXiv:hep-th/9301111];
  R.~Chatterjee and L.~Takhtajan,
  Lett.\ Math.\ Phys.\  {\bf 37}, 475 (1996)
  [arXiv:hep-th/9507125].
  
  
  \bibitem{triples}
H.~Awata, M.~Li, D.~Minic and T.~Yoneya,
  JHEP {\bf 0102}, 013 (2001)
  [arXiv:hep-th/9906248] and references therein.
  
  
\bibitem{vector}
R. Leigh, A. Mauri, D. Minic and T. Petkou, 
\ Phys.\ Rev.\ Lett.  {\bf 104}, 221801 (2010) [arXiv:1002:2437]. See also, 
J. Bagger and N. Lambert, Phys. Rev. {\bf D75}, 045020
(2007); Phys. Rev. {\bf D77}, 065008 (2008); JHEP {\bf 0802}, 105
(2008); A. Gustavsson, Nucl. Phys. {\bf B811}, 66 (2009)
and references therein.

\bibitem{zakharov}
For a discussion of turbulence at weak coupling, consult
V.E. Zakharov, V.S. Lvov and G. Falkovich,
{\it Kolmogorov spectra of turbulence}, Springer, 1992.
  

\bibitem{migdal}
A. A. Migdal, arXiv:hep-th/9303130; Int. J. Mod. Phys. {\bf A9}, 1197 (1994)
[arXiv:hep-th/9310088]; arXiv:hep-th/9306152.

\bibitem{wilson}
K. Wilson, Phys. Rev. {\bf D10}, 2445 (1974).

\bibitem{staples}
These numerical studies aim at determining whether or not the correct scaling for
$\langle v_i(l) v_j (0) \rangle$ results, given a certain turbulent velocity field 
with a certain energy spectrum scaling. Interestingly, the preliminary results from
the numerical simulations nearly exactly reproduce the correct scaling for
$\langle v_i(l) v_j (0) \rangle$ when given a velocity field with a Kolmogorov
energy spectrum [A. E. Staples, work in progress].

\bibitem{foot1}
Whereas this is the usual picture, the situation is reversed if the forcing occurs at high and low 
mesoscale (inertial range) wavenumbers, and not at a single intermediate mesoscale wavenumber 
location. See, e.g., D. K. Lilly, J. Atmos. Sci. {\bf 46}, 2026 (1989).
  
\end{thebibliography}
\end{document}